\documentclass[10pt]{iopart}

\usepackage[table]{xcolor}
\usepackage{caption}
\definecolor{Gr}{gray}{0.9}
\usepackage{bm}
\usepackage{graphicx,stackengine}
\usepackage[percent]{overpic}
\usepackage{amsfonts}
\usepackage{hyperref}
\hypersetup{
    colorlinks=true,
    linkcolor=blue,
    filecolor=magenta,      
    urlcolor=cyan,
    }
\usepackage{xcolor}

\usepackage{multirow}

\usepackage{tikz}
\usepackage{tikz-qtree}
\usetikzlibrary{trees}	
\usepackage{amsmath} 
\usepackage{kantlipsum,widetext}

\begin{document}
\title[Hallucinations in medical devices]{Hallucinations in medical devices}
\author{Jason Granstedt, Prabhat Kc, Rucha Deshpande, Victor Garcia, Aldo Badano}

\address{
Division of Imaging, Diagnostics, and Software Reliability, Office of Science and Engineering Laboratories, Center for Devices and Radiological Health, \\U. S. Food and Drug Administration, Silver Spring, MD 20993}
\ead{jason.granstedt@fda.hhs.gov}
\vspace{10pt}
\begin{indented}
\item[]\today
\end{indented}

\begin{abstract}
Computer methods in medical devices are frequently imperfect and are known to produce errors in clinical or diagnostic tasks. However, when deep learning and data-based approaches yield output that exhibit errors, the devices are frequently said to hallucinate. 
Drawing from theoretical developments and empirical studies in multiple medical device areas,
we introduce a practical and universal definition that denotes hallucinations as a type of error that is plausible and can be either impactful or benign to the task at hand. The definition aims at facilitating the evaluation of medical devices that suffer from hallucinations across product areas. Using examples from imaging and non-imaging applications, we explore how the proposed definition relates to evaluation methodologies and discuss existing approaches for minimizing the prevalence of hallucinations.
\end{abstract}
\vspace{2pc}
\noindent{\it Keywords}: hallucinations; deep learning; artificial intelligence; generative models; medical imaging
\submitto{Artificial Intelligence in the Life Sciences}
\maketitle

\tableofcontents 

\section{Introduction}

The phenomenon of hallucinations within AI systems can adversely affect the efficacy of algorithmic applications by diminishing user trust and introducing safety hazards in critical contexts. In other contexts, hallucinations may offer advantages, such as in the creation of innovative content or the production of synthetic data for model training. Hallucinations pose substantial challenges particularly in high-stakes applications where accuracy is imperative. Within AI applications in medical devices, hallucinations may influence clinical decision-making and potentially jeopardize patient outcomes through diagnostic or therapeutic errors. 
Despite the concept of hallucination having been introduced to the scholarly community about a decade ago, a definitive and universally recognized definition pertaining to hallucinations in medical devices is currently absent. This article delineates an approach designed to provide a clear context for referring to hallucinations in outputs of AI medical applications, thereby aiding in the assessment and prevention of such phenomena within the methodologies for medical device evaluation.

The notion is often articulated through the term `confabulation,' defined as `the generation of confidently asserted yet erroneous or misleading content by which users may be misled or deceived.' 
Recently, Xu et al. have offered a more pragmatic approach to defining hallucinations with a theoretical framework in which hallucinations are delineated as the discrepancies between generated outputs and a ground truth function. By leveraging learning theory, they elucidate that hallucination is inherently unavoidable and that the complete eradication of hallucinations from real-world large language models (LLMs) is not feasible.

\begin{figure}
    \centering
    \includegraphics[width=0.9\linewidth]{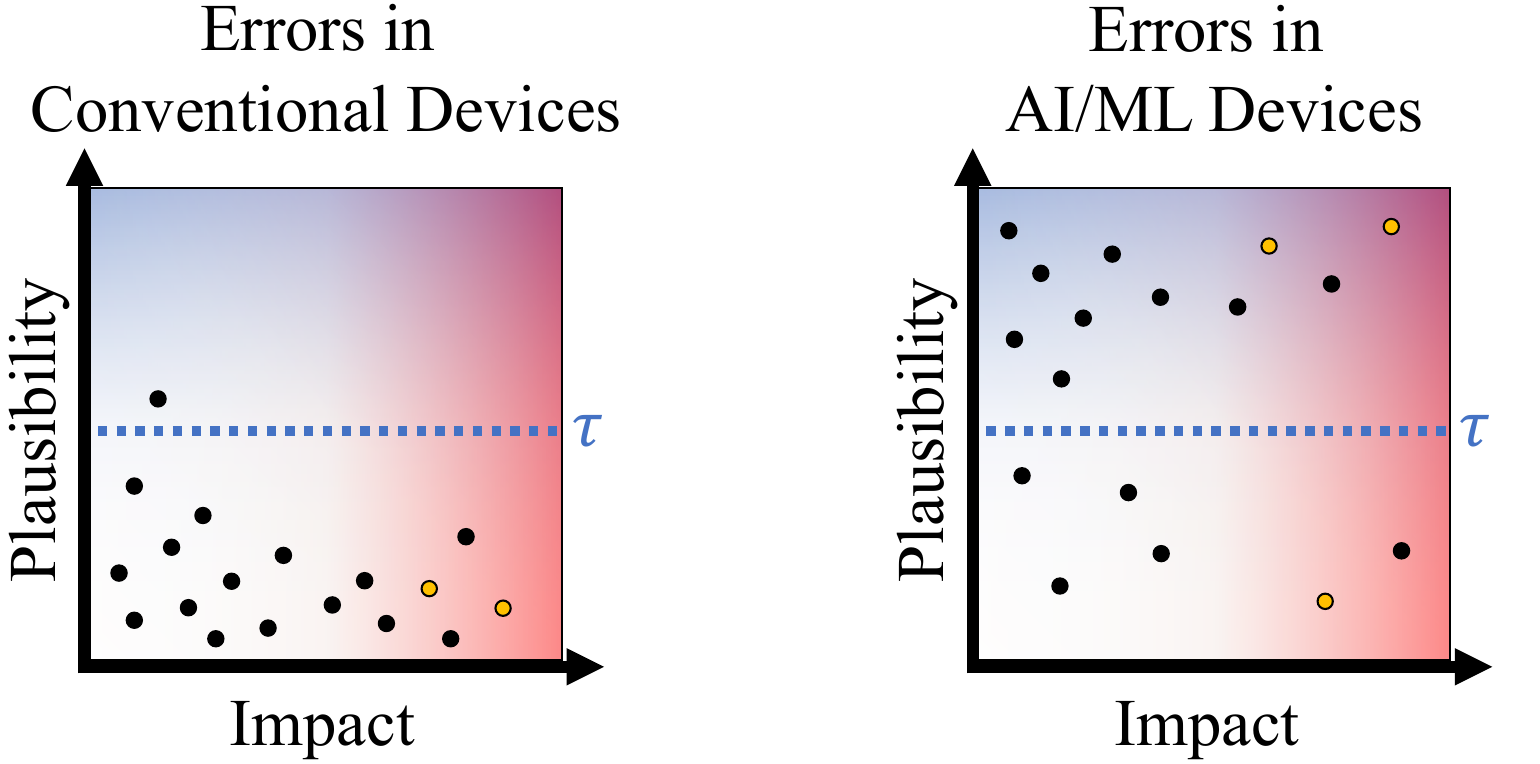}
    \caption{Mock diagram of errors from a conventional and AI-enabled device, plotted against axes of impact and plausibility. Unmitigated impactful errors are colored yellow. Plausibility introduces another risk vector, as such errors may lie outside the domain of conventional risk mitigation strategies and clinician intuition. Thus, an AI model may lead to worse patient outcomes even if it produces fewer impactful errors, as the plausibility of such errors may circumvent the traditional guardrails of medical devices. Errors above a certain plausibility threshold $\tau$ are labeled as hallucinations per our definition to identify such risks.}
    \label{fig:mock_ip}
\end{figure}

Expanding upon Xu's framework, we propose to define hallucinations as a subset of errors. Hallucinations are identified as plausible errors with two distinct subtypes: (1) impactful hallucinations and (2) benign hallucinations. Impactful hallucinations negatively impact device performance, whereas benign hallucinations have no significant effect. Additionally, there exist non-hallucination errors, which are characterized by their obviousness and traceability to device artifacts or pre-specified failure modes. To determine whether an error is a hallucination or a non-hallucination error, we consider the nature of both the assessment task and device user. This definition does not specify the type of user, and the determination of whether an error is plausible or subtle is contingent upon the nature and level of expertise of the user which, according to the intent of the evaluation framework, could be a domain expert, a naive user, or in some cases, an algorithmic interpreter. This approach to defining hallucinations is consistent with other work by~\cite{wei2024measuring} where hallucinations are defined as ``false outputs or answers that are not substantiated by evidence'', which is equivalent to Xu's definition linked to ground truth functions in certain cases. A diagram of the considered axes for our definition is included in Figure~\ref{fig:mock_ip}. 

The primary departure from our definition in other proposals is the incorporation of plausibility. 
Plausibility is a continuum and the threshold at which an error becomes sufficiently plausible to be labeled a hallucination is likely observer and task-specific. 
It may be possible to conduct studies to determine this threshold for relevant use cases but such approaches are outside of the scope of this work. 
Nonetheless, the role of plausibility (and the different proficiencies of observers) has been previously established in the medical domain. 
A 2013 Pew survey revealed that doctors disagreed with patients' self-diagnoses informed by online resources approximately a third of the time~\cite{pew2013health}. 
Such resources provided answers that were plausible to the patient, but were readily discernible as incorrect by a medical professional. 
The danger with AI methods is that errors may be so plausible that they may fool even experts. 
There are several recent occurrences of such errors in the legal profession~\cite{case2023matavavianca, case2025kovli, case2025lacyvstatefarm}; it is likely that health professionals will also be susceptible. 
These hallucinations herald a new risk vector that can circumvent professional intuition and bypass current risk mitigation strategies. 

An additional departure from previous definitions is our introduction of the concepts of benign and impactful hallucinations. While other work focuses on the taxonomy of a hallucination, what is more important in the medical field is the impact on patient care.
Human judgment has been demonstrated to be susceptible to AI errors~\cite{agudo2024impact, nagendran2023impact, jacobs2021machine} and clinicians can inherit AI model biases~\cite{vicente2023humans}. A complicating factor is that what may originally be perceived as a benign hallucination may become impactful if the information is later used to affect patient care decisions. If the output may be referred to later, then there is always a risk of downstream error propagation. 

Due to the high-risk nature of medical applications, tolerance for hallucinations is low. 
Minor modifications to factual details can impact patient management, so systems that produce hallucinations may degrade clinician trust and decrease utility even when the hallucination is benign~\cite{kim2025medical}. Every poor-quality system deployed further degrades trust in AI as a whole and leads to an increasing skepticism towards future applications. 

While there are many types of AI-enabled medical devices, in this work we will focus on three areas: imaging devices, generative-based synthetic medical images, and large language models. These three areas have seen explosive growth and extensive implementation of the types of AI models that lead to the proliferation of hallucinations. We will begin by describing the types of hallucinations in these device types in Section~\ref{sec:hmd} and then discuss how these hallucinations may be quantified and mitigated in Sections~\ref{sec:mmh} and~\ref{sec:mh}, respectively. Finally, we will conclude with a summary of the impact of hallucinations on the future of AI-enabled medical device development in Section~\ref{sec:summary}.

\section{Hallucinations in Medical Devices}
\label{sec:hmd}

\subsection{Imaging devices}
\label{sec:imgdev}
Inverse problems in imaging have been an 
active area of investigation for AI methods~\cite{arridge2019solving, liang2020deep, mccann2017convolutional, ongie2020deep, wang2020deep, reader2020deep, wang2018image}.
A critical task in many applications in medical, scientific, and industrial applications is the recovery of an image from a set of measurements, which are frequently noisy and incomplete.
Improving the quantity or quality of these measurements often has an associated cost. 
Thus, it can be desirable to explore computational techniques to improve the utility of the image. 
One such method is regularization, which encodes desired attributes for an image into a mathematical formula that is applied during the reconstruction process to recover a more applicable image. 

Recent research works have been predominantly focused on strategies that learn a prior distribution from a dataset via neural networks.
Instead of a handcrafted term employed by conventional regularization strategies~\cite{smith2016null, hahn2016null, kelly2017deep, schwab2019deep, rowbotham1997improved, deal1996nullspace, ongie2020deep}, these methods are data-driven.
Though these methods have achieved state-of-the-art results in several areas~\cite{wang2016perspective, mccann2017convolutional, ravishankar2020image}, there are rising concerns about generalization performance and robustness~\cite{gottschling2020troublesome, antun2020instabilities}.
This is of particular concern in medical imaging, where even a large dataset may lack rare abnormalities.

These stability issues sometimes result in the generation of false structures in reconstructed images~\cite{belthangady2019applications, hoffman2021promise, varoquaux2022machine}, which have been referred to as hallucinations.
Studies have warned of the potential for misdiagnoses from these hallucinated structures~\cite{knoll2020advancing, muckley2021results}.
These concerns have recently been validated with clinical samples~\cite{bosbach2024deep}. 

The robustness of neural networks has been investigated in many fields~\cite{eykholt2018robust, carlini2018audio, finlayson2019adversarial, heaven2019deep, bastounis2021mathematics}. 
Some of these approaches consider worst-case small permutations to the input of the network~\cite{huang2018some, antun2020instabilities, darestani2021measuring, genzel2022solving}, while others consider alternative adversarial methods~\cite{raj2020improving, morshuis2022adversarial, alaifari2023localized}. 
A recently developed tool for analyzing neural network reconstructions for these phenomena is hallucination maps, which allow the isolation of artifacts associated with imperfect priors~\cite{bhadra2021hallucinations}. 
Various approaches have been proposed to incorporate information about an imaging system into neural network reconstruction methods and demonstrated resilience to these adversarial approaches~\cite{schwab2019deep, gottschling2020troublesome, colbrook2022difficulty}. 
Adding noise to the dataset has also demonstrated to be effective at increasing stability, albeit at the expense of performance~\cite{gottschling2020troublesome}. 

Regularization techniques can improve human observer performance, but they cannot add any additional information to a reconstruction~\cite{zhang2021impact}.
This is a fundamental limitation of imaging systems - information is always lost during the imaging process, and no post-processing can recover diagnostic details if a device does not measure the relevant details~\cite{wagner1985unified, zhang2021impact, bhadra2021hallucinations}.
Effectively, the relative increase in image quality comes with the trade-off of instability and the resulting hallucinations~\cite{gottschling2020troublesome}.
This is an inherent flaw of data-driven approaches.

Exploring hallucinations in imaging devices presents a unique opportunity due to the accessibility of ground truth that can lead to the certain identification of errors.
Nevertheless, what makes an error "plausible" remains frustratingly elusive.
Plausibility can vary depending on the downstream task and whether the image is employed by a human or an algorithm. 
There can also be significant differences in plausibility between humans, based on level of training or simply sheer variability. 
Nonetheless, plausibility is one of the most complex aspects of hallucinations in a clinical environment.
Neural network reconstructions can lead to an overestimation of the diagnostic utility of an image disconnected from the quality of the underlying measurements, which can subvert the intuition of a reader~\cite{knoll2020advancing, muckley2021results}.

The predictable behavior of conventional regularizers enables clinicians to recognize and adapt to the various errors that arise from their use. For instance, a radiologist can turn off an image enhancement-based smoothing option in a radiological image acquisition system if the radiologist deems that a lesion in the acquired radiological image has been over-smoothed. 

Task-based evaluation through reader studies, both human and computational, is one method for evaluating performance on downstream tasks~\cite{vennart1997icru, wagner1985unified, barrett1993model}.
However, images are sometimes employed for multiple tasks and improvements in one area may come at the expense of another.
Some new datasets have begun to bridge this gap by providing diagnostic information~\cite{zhao2022fastmri+}, but access to larger datasets and deployment of multi-task evaluations will likely be necessary to assess the utility of neural network reconstructions. 

The driving force in the technological advancement of medical imaging has been less radiation\footnote{As Low as Reasonably Achievable (ALARA) has been the guiding principle of radiation safety when using imaging modalities like CT. ALARA advocates for dose optimization while maintaining the image quality required to perform the diagnostic task at hand accurately. As such, increasing the dose level – for large patients – would be consistent with ALARA's principle \cite{cynthia_ct_dose_strategy}.} and saving scan time in the last two decades. AI-based methods are being proposed to supplant conventional physics-based methods (like the Filtered BackProjection \cite{kak_fbp_chap} and inverse Fourier transform \cite{mri_ifft_book_chap}) such that one can faithfully recover internal organs corresponding to the person using measurements acquired at very low-dose \cite{chen_redcnn_ct_paper,wgan_ldct} or under-sampled rate \cite{unrolled_deep_admm_net}. Recently, domain-transfer-based applications have also been proposed such that images acquired using a given modality (or a sub-modality) can be seamlessly transformed into a different modality (or sub-modality) \cite{med_analysis_cyclegan_chap}. For instance, cycleGAN has been proposed to translate an MRI image to its CT counterpart \cite{cyclegan_mr_2_ct}. However, due to the ``curse of data processing inequality'' \cite{dpi_theory}, an AI-based method might compensate for the information lost due to hardware-based less radiation, undersampled-acquisition, or lack of the imaging domain-specific properties with the data priors that are not specific to the person being scanned \cite{varun_hallu, tam2012null}. 
Simply put, as measurement quality deteriorates, AI models become more unstable\footnote{A model is unstable when a small perturbation in input to the model leads to a large fluctuation in the model output.}. This subsequently leads to imaging errors \cite{cyclegan_mr_hallu} that cannot be distinguished as conventional artifacts, either in terms of their obviousness from our past use of imaging devices or their traceability to imaging system-based shortcomings. We refer to them as hallucinations.

\begin{figure*}[!bt]
\centering
\includegraphics[width=1.0\linewidth]{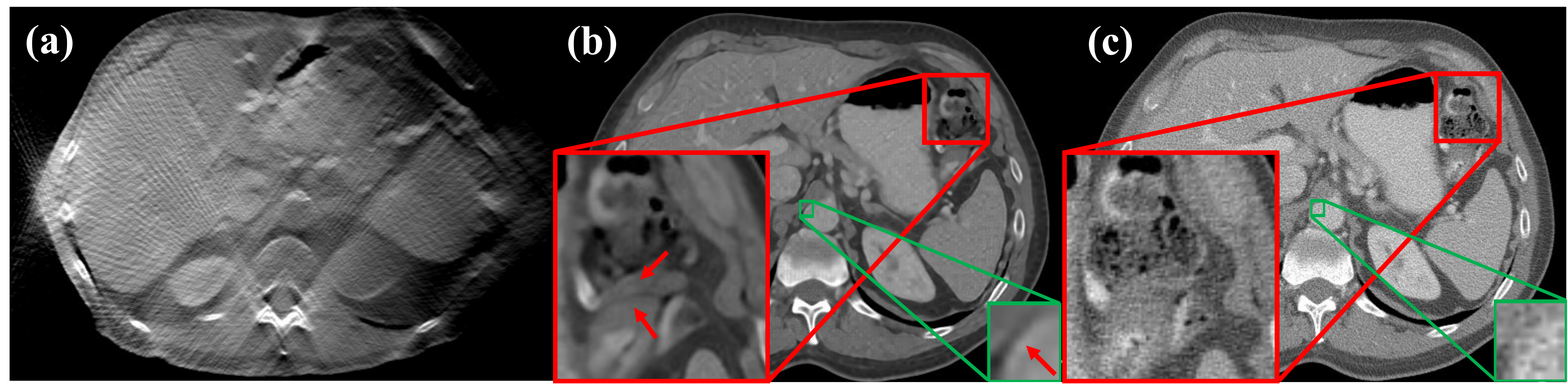}
\caption{An illustration of artifacts that are readily discernible in (a) and non-discernible in (b) to human eyes. Hallucinations (related to the untrue addition of the two loops of bowels and plaque-like features indicated by the red arrows) in the AI-based output in (b) only become evident after comparison against its reference image in (c). The CT image in (c) is obtained by applying a physics-based analytical algorithm on its fully sampled raw measurements. The one in (b) is obtained by applying an AI-based super-resolution model on the four times downsampled version of (c).}
\label{img:mw_n_ai_hallu}
\end{figure*}

An essential hallmark of hallucinations in medical imaging is that – unlike conventional artifacts such as distortions, line artifacts, beam hardening, gibbs ringing, aliasing, etc. \cite{ct_artifact_chap,mri_artifacts}– it may not be possible to identify all the hallucinated features without the corresponding reference image. This phenomenon is depicted in fig.\,\ref{img:mw_n_ai_hallu}(b). Only in the presence of a reference image (fig.\,\ref{img:mw_n_ai_hallu}(c))– acquired using complete data and physics-based method – and a thorough review of AI-based reconstruction (fig.\,\ref{img:mw_n_ai_hallu}(b)) do all the factually incorrect features reconstructed by the AI become evident. In contrast, the line artifacts in fig.\,\ref{img:mw_n_ai_hallu}(a) are readily discernible to human eyes and can be traced to the limited angular tilt of the imaging system when acquiring the data. Hence, per our definition, the CT image in fig.\,\ref{img:mw_n_ai_hallu}(a) would constitute a non-hallucinatory or conventional artifact while fig.\,\ref{img:mw_n_ai_hallu}(b) would constitute a hallucination. Further, both images in the figs.\,\ref{img:mw_n_ai_hallu}(a, b), would qualify as impactful errors. 

From the perspective of information theory, a typical $512\times512$ image encodes much more information than a single page with $500$ words. 
A typical medical imaging-based denoising or reconstruction problem incorporates the raw data acquired from a patient (or is a conditional problem; more information in section \ref{sec:condAI}). Hence, a large number of hallucinations in the denoising and reconstruction domain may be more subtle and impactful than nonsensical or benign compared to what we may observe in language-based or unconditional domains. However, the nature of impactful versus benign hallucination for the imaging problem may also vary based upon the nature of challenging (ill-conditioned \cite{per_just_enough}) problems one seeks to resolve. For instance, consider a case whereby only half of a patient's internal body part is scanned and AI is used to predict the remaining half. 
This might yield highly perturbed/nonsensical outputs that experts may easily be able to categorize as errors. Overall, AI in medical imaging may yield a range of errors that may be subtle to obvious and may have impactful to benign harm. As such, it is critically important to use benchmarked imaging datasets (with patient-based diseased labels from patient follow-up data) \cite{fastmri_plus,armato2016lungx,yan2018deeplesion} and perform various downstream evaluations \cite{udandarao2024no} (such as pathology-based classification, quantification, detection, discrimination, etc.) to understand the nature of AI hallucinations for a given imaging application.

\subsection{Synthetic data} 
Generative AI for data augmentation holds great promise for learning-based methods in the medical domain as it may address data scarcity issues while maintaining patient privacy. It has been employed for generating synthetic data in various imaging modalities \cite{chen2022generative, garcea2023data} such as ultrasound imaging \cite{xu2024synthetic}, mammography \cite{lee2022analyzing} and histopathology \cite{xue2021selective}. Typically, in data augmentation applications, the generation task is one of two kinds: 1) ``unconditional generation" or generation with no prompts (only random noise as an input), e.g, given a dataset of chest radiographs, generate a similar synthetic dataset, and 2) conditional generation where the prompt may be a class-label, feature value, or another image, e.g., generate a mammogram from a given breast type (class-label), or generate a T2-weighted medical resonance (MR) image given the corresponding T1-weighted MR image of the same patient. Although conditional generation may ensure consistency with the input condition (for a well-trained generative model), it does not preclude hallucinations in features that are uncorrelated with the conditioning input. 

\subsubsection{Unconditional generation} 
A distinctive aspect of unprompted/unconditional generation of images is that each generated image is entirely synthetic and does not correspond to any individual in the real world. These synthetic images can still have defined ground truth functions and hallucinations, but the ``hallucination is no longer related to correctness or factualness in the real world''  \cite{xu2024hallucination}. Specifically, ground truth functions describe the anatomical knowledge represented in the \emph{entire} training data and can be considered as a mapping between hidden variables and images in the training set. 
In unconditional generation, an AI model generates new content by seeking to learn the underlying patterns of the training data without receiving explicit guidance, human labels, instructions, or a priori constraints. Inconsistencies or errors with respect to the ground truth function might still exist if the generative model function fails to learn the ground truth function. These inconsistencies may manifest as network artifacts and/or hallucinations. Recall that the difference between the two is highly subjective and based on perceptual plausibility and that lower plausibility does \emph{not} necessarily imply lower downstream clinical impact.

In literature, hallucinations have been reported in various attributes such as per-image feature prevalence, feature-specific intensity distributions, and relative feature locations, both in domain-agnostic \cite{deshpande2024method, deshpande2024assessing} and domain-specific studies \cite{kelkar2023assessing}. Furthermore, some works report network artifacts and hallucinations under the same terminology and both are commonly known to occur in generative tasks of images \cite{muller2023multimodal, deshpande2025report} in practice. Some examples of hallucinations reported in literature according to the proposed definition are multiple optical disks instead of one in eye fundus images (as expected from the training data) and unexpected locations of medical devices in chest radiographs \cite{muller2023multimodal}. Examples of network artifacts include checkerboard artifacts in histopathology images \cite{muller2023multimodal} and nipple artifacts in mammography images \cite{lee2022analyzing, deshpande2025knowledge}. 

\subsubsection{Conditional generation}\label{sec:condAI} In prompted or conditional image generation, the generated image may be 1) entirely synthetic (e.g., when the prompt is a class-label) or 2) partially synthetic (e.g., when a patient image in one imaging modality is to be transformed to another), i.e., a domain transfer task. In the first case, the ground truth function and hallucinations are defined similarly to unconditional generation, only with the assumption that the generative model function will be consistent, i.e., not hallucinate with respect to the conditioning input and correlated attributes. 

In the second case (domain transfer task), when assumptions of data sufficiency and relevance are met, a unique ground truth function may be computable from the training data. Here, the ground truth function encompasses logical consistencies and relative anatomical mappings between domains, which can intuitively be understood as a bijective mapping between the domains. If the generative model function fails to learn this ground truth function, the resulting inconsistencies or errors between the two will lead to hallucinations. 

However, the ground truth function may not be computable if the training dataset does not contain relevant and sufficient information for the generation task. In that case, hallucinations will occur (assumptions for definition 4 in \cite{xu2024hallucination}). One scenario when the ground truth function is not uniquely computable is when the physics of the input and output domains differs vastly for a given anatomy. For example, in a generation task where computed tomography (CT) is to be generated from positron emission tomography (PET) image inputs, hallucinations \emph{must be expected} in the generated images as a unique ground truth mapping cannot be computed from the training data for the two imaging domains. Thus, the use of generative models in such problems is not advisable. Similarly, when the dataset is not relevant for a generation task, e.g., domain transfer of a diseased patient when the disease case was absent in the training data (detailed demonstration in \cite{cohen2018distribution}), a ground truth function does not exist and hallucinations must be expected. 

Examples of hallucination in conditional generation tasks in literature include: addition of tumors in T1 MRI that did not exist in FLAIR MRI \cite{cohen2018distribution}, unexpected addition of realistic histopathological features \cite{vasiljevic2022cyclegan} in virtual staining.

In both unconditional and conditional generation, errors in the generative model function may arise from various factors such as: (i) ineffective latent encoding\cite{bond2021deep}, (ii) distribution-matching loss functions \cite{cohen2018distribution}, and (iii) insufficient receptive field in the network architecture \cite{vasiljevic2022cyclegan}. As generative models continue to evolve, so do the manifestations of their hallucinations and network artifacts.

\subsection{Language and multimodal devices}
Large language models have been applied to many natural language tasks, from text summarization~\cite{zhang2024benchmarking} to question answering systems~\cite{tan2023evaluation} and machine translation~\cite{zhu2023multilingual}. 
An interesting property of these tasks is the relative sensitivity to errors based on the application - while errors in summarization quickly become impactful due to the concern of fidelity~\cite{pagnoni2021understanding}, question answering systems may permit more errors to achieve the secondary objective of user engagement~\cite{huang2020challenges}. 
The medical versions of these tasks are likewise varied, which leads to a corresponding spread in health risks; a model for generating radiology impressions has a notably different risk profile than one performing physician note summarization.

Many of the errors that LLMs make can appear to be plausible, in part due to the ability of LLMs to correctly mimic grammar. 
For instance, a patient record summarization summary with an erroneously inserted diagnosis is likely to be plausible to all but a doctor who is intimately familiar with that patient's medical history. 
It is important to remember, however, that LLMs are trained to produce the most likely outputs, not the most accurate ones~\cite{radford2019language}.
Indeed, it can be argued that the outputs of such models are persuasive  because they have been specifically trained to produce plausible answers to convince humans~\cite{ouyang2022training}. 
While the likelihood of answers can be correlated to truth~\cite{radford2019language, devlin2019bert, roberts2020much}, the two are not the same.
This issue becomes especially prevalent in long-tailed domains where low probability events are critical for understanding the complex systems involved~\cite{taleb2007black}, such as law and medicine.
One could consider truthful, accurate, and modern data itself as inherently long-tailed~\cite{ji2023survey}. 

A complication with LLMs is that for any given task, there is usually more than one output that satisfies the query~\cite{su2020diversifying, guan2020union}. Human evaluation~\cite{shuster2021retrieval, tu2024conversational, tu2024generalist, singhal2025qa, li2025holistic, tanno2025collaboration} continues to remain as the preferred means of developing a reference standard for LLMs because humans can assess the various application of LLMs, including objective and subjective evaluation metrics. Unfortunately, generating a reference standard using humans is resource intensive, costing both time and money. Many public datasets have been created to evaluate LLMs for pre-specified tasks~\cite{wang2018glue, liang2022holistic, hendrycks2020measuring}, and newer evaluation approaches of an LLM-as-a-Judge framework leverage LLMs to replace human assessments~\cite{zheng2023judging}. However, such methods are limited in reasoning capability and it is unclear if they will generalize to long-tailed domains such as medicine. 

There are various approaches for increasing the utility of LLMs for particular applications~\cite{patil2024review}.
In reinforcement learning techniques, humans analyze model responses and indicate preferential answers~\cite{ouyang2022training, touvron2023llama}. Models may also be fine-tuned trained on specialized datasets to impart knowledge on a particular area~\cite{tu2024generalist} and to implement safety guardrails~\cite{touvron2023llama}. However, one challenge with fine-tuning approaches for models is that model performance may improve in one area but degrade in another potentially relevant areas; this is referred to as an alignment tax~\cite{lin2023mitigating}. In deployed models that continually learn, this performance trade-off may occur unintentionally during retraining leading to unintended performance drift. Additionally, fine tuning on additional training samples may remove fine-tuned weights~\cite{shen2024anything}, even when such a result is unintentional~\cite{qi2023fine}.  

Assembling a robust dataset for medical tasks for such fine-tuning is also frequently not trivial. Ensuring a clean dataset is also essential for model performance~\cite{zhang2023siren, penedo2023refinedweb}. For foundation models, this can come both in upsampling high-quality data during pretraining~\cite{touvron2023llama} or fine-tuning to a specific task~\cite{singhal2025toward, chung2024scaling, wu2024pmc}. The ground truth can be unknown due to patient drop-out, lack of follow-up data on disease-based outcomes at the patient level, or disagreements among doctors.
Additionally, even the data employed to train the models may be suspect.
Models may rely on documentation that is either misleading, contains wrongly imputed/extrapolated/interpolated labels when compensating for missing data, or is out of date.

In addition to LLMs, visual language models (VLMs) are now being explored for applications in the medical domain~\cite{hartsock2024vision}. While these models can generate much more detailed responses to visual inputs than previous visual question answering systems, this comes with a corresponding increase in hallucination risks. At the basic level, such models have demonstrated vulnerability to object hallucination, i.e., a description from VLM that is inconsistent with the target image~\cite{li2023evaluating}. Furthermore, commonly-used representations for the language-visual alignment training have demonstrated shortcomings in representations for object counts, viewpoints, and orientations~\cite{tong2024eyes}. Finally, errors in the alignment between the vision and language can result in an observable gap between the visual backbone of the model and the visual recognition capabilities of the LLM~\cite{zhai2023investigating}.

\section{Quantifying Hallucinations}
\label{sec:mmh}

The determination of an artifact as a hallucination is difficult to formally define, as it relies on the artifact being ``plausible''. Thus, what may appear as an obvious error to one observer (human or mathematical) may instead fool another. The severity of hallucinations varies a great deal depending on context. Minor artifacts in irrelevant portions of an output seldom have an effect on plans of care, but the same distortion in another location can lead to a misdiagnosis. 

Thus, a hallucination can be impactful for one task and benign for another, and may be plausible for one observer (such as a patient) and obvious to another (such as a doctor). As a result, there is a great deal of subjectivity inherent in identifying hallucinations relevant to clinical care, and hallucinations are highly application and user dependent. Nevertheless, some algorithmic approaches have been proposed to quantify and mitigate hallucinations. 

One connection that has been made across multiple fields is that hallucinations are connected to stability. 
While the implementation may differ, the core concept is the same - AI devices that produce substantially different outputs from a small perturbation in the input are more likely to produce hallucinations.
This connection was first made in imaging, where a trade-off was observed between global metrics (e.g., mean squared error) and stability~\cite{gottschling2020troublesome}.
This observation has more recently been repeated in image generation~\cite{tivnan2024hallucination} and LLMs~\cite{mohri2024language, cherian2024large}.
Many developed methods exploit this relationship to measure and adjust the trade-offs between hallucinations and performance.

When the ground truth is known, as in many inverse problem simulations, a straightforward method for measuring stability is worst-case small permutations~\cite{antun2020instabilities}. 
The input is modified by a small amount and iteratively optimized to maximize the change in the output. 
However, while this method is sufficient for demonstrating that neural network reconstruction techniques are unstable~\cite{gottschling2020troublesome}, it does not provide the necessary measure of plausibility to evaluate if the alteration is a hallucination. Hence, as previously mentioned in the section \ref{sec:imgdev}, rather than relying on anecdotal accounts of the efficacy of a new AI reconstruction model, a preferred methodology is to objectively quantify the error of the model for imaging tasks such as quantification, detection, discrimination, or prediction to assess the impact of hallucinations.

For generative tasks in both imaging and language, the ground truth is less defined. To measure hallucinations in these domains, special datasets have been constructed.
For generative models, these take the form of purposefully created stochastic models that encode attributes of interest in medical imaging~\cite{deshpande2024method, deshpande2024assessing}.
However, while such methods are valuable for demonstrating relative performance of generative models on the specific task, it remains unknown how the results generalize to the broader medical imaging domain.
When the ground truth is accessible, such as in a generative reconstruction task, another method of evaluating hallucinations is the hallucination index~\cite{tivnan2024hallucination}. This method computes the Hellinger distance between the distributions of the ground truth and the reconstructed images. Still, even after such computation, determining the cut-off that dichotomizes faithful and hallucinated reconstruction may not be trivial.

For evaluating LLM performance, datasets have been constructed to specifically probe for hallucinations in the medical domain~\cite{pal2023med, xia2024cares, chen2024detecting}. 
However, LLMs have been demonstrated to be highly unstable with performance metrics varying dramatically depending on the instruction set~\cite{mizrahi2024state} and permutations in the input prompt~\cite{wang2021adversarial, wang2023robustness, wang2023decodingtrust}.
This results in the LLMs not performing as well in practice as the obtained metrics would suggest.
Thus, understanding the impact of this stability is crucial to understand how an LLM may perform in a medical environment.

\section{Minimizing Hallucinations}
\label{sec:mh}
    As with hallucination detection, similar concepts have been applied across fields to reduce hallucinations. 
    Such methods involve incorporating either a measure of truth or a concept of uncertainty into the training or evaluation method.

    Truth in imaging is governed by the acquired measurements and the properties of the imaging system. 
    One such approach in imaging is the null-space shuttle method~\cite{deal1996nullspace, schwab2019deep}.
    This modification prevents the neural network from modifying the information in the measurement data and only permits reconstruction on unknown parameters of the image.
    Thus, it can prevent hallucinations from the captured data.
    However, the risk of hallucinations remains for components of the image that are measured.
    Furthermore, some modifications to the measurement data, such as mitigating measurement noise, are beneficial.
    The null space shuttle procedure prevents the network from learning these tasks, necessitating the inclusion of additional elements in the pipeline if such features are desired.

    To permit the learning of such features, other network architectures for reconstruction consider softer constraints~\cite{hammernik2018learning, chen2018learn}. 
    Thus, data fidelity is incorporated but it is not as binding as null-space shuttle procedures. While this method may reduce hallucinations by mitigating some divergences from the ground truth, it remains unknown if the trade-off is worthwhile in many medical imaging cases.
    An alternative approach instead considers incorporating uncertainty into the measurements by injecting noise during the training process~\cite{gottschling2020troublesome}.
    This method is able to improve the stability of neural network methods, but comes with a corresponding reduction in performance.

    For generative imaging applications, a corresponding implementation is the AmbientGAN~\cite{bora2018ambientgan}. This modification to the typical GAN framework includes a measurement function that encodes information about an imaging system. 

    Other methods are useful for LLMs to incorporate relevant knowledge for queries.
    Retrieval augmented generation (RAG) is one of these techniques, which searches for relevant documents to append to the query~\cite{lee2019latent, guu2020retrieval, lewis2020retrieval}.
    However, RAG remains vulnerable to many hallucination vectors.
    In many modern implementations, an LLM is responsible for generating the RAG call from the original query. This process remains susceptible to some of the baseline hallucinations observed in LLMs.
    Additionally, retrieving substandard or out of date evidence may degrade model performance further~\cite{xu2024knowledge}.
    Finally, such methods are also susceptible to the alignment tax.
    Knowledge graphs are also being explored for hallucination detection by providing explicit facts and reasoning~\cite{lavrinovics2025knowledge}, but likewise suffer from many of the above issues.

    Another way of mitigating hallucinations in LLMs is post-processing of the response to remove potentially erroneous information. Some proposed method employ conformal probability to redact portions of the LLM's response~\cite{mohri2024language, cherian2024large}. However, to obtain a high confidence of factual information, such methods frequently prune substantial amounts from the response which potentially limits the utility of the method.
    
    Finally, many prompt-engineering based strategies have been employed in an attempt to reduce hallucinations.
    Some approaches prompt a chain of reasoning for the models~\cite{ji2023towards}.
    Others use ensemble methods, either with a collection of models~\cite{du2023improving} or a single model with multiple personas~\cite{wang2023unleashing}, to produce and analyze multiple responses.
    In many ways, such approaches are reminiscent of averaging across several random samples to increase confidence in the output.
    However, it has been convincingly argued that such approaches are unable to prevent hallucinations~\cite{xu2024hallucination}.

    While many of the methods discussed in this section may improve performance on physical metrics or even tasks, none of them prevent hallucinations from occurring. Thus, there always remains a risk when employing AI for medical tasks.

\section{Summary}
\label{sec:summary}
Medical devices are employed in many applications that impact patient care, both directly and indirectly. 
The incorporation of AI/ML methods into these devices contains both benefits and risks. It is important to emphasize that AI models in devices do not need to be perfect to be useful, especially when such models demonstrate performance improvements over existing standards of care. 
However, hallucinations pose novel challenges to the existing medical ecosystem.
By focusing the discussion of hallucinations on downstream impacts to patient care, meaningful progress can be made for the safe and effective integration of AI-enabled medical devices. 

Nonetheless,  hallucinations can not be fully removed as they are intrinsic to neural network-based methods and attempts to reduce hallucinations may come at the cost of decreased performance. 
This inherent instability introduces unique risks due to injecting errors in the chain of care that may not manifest until much later. 
Furthermore, the risk belongs to the entire stack of deployed AI models. 
While each model may be individually low-risk, the combined system may become high-risk due to cascading instabilities.
Finally, integration of AI devices into clinical workflows can result in decreased clinical efficiency and rendering medical databases unsuitable for training future generations of AI algorithms.

\section*{References}
\bibliographystyle{ieeetr}   %
\bibliography{bibliography}
\clearpage
\end{document}